\hoffset = +2mm
\baselineskip=6mm
\vglue1cm 
\line
{\vbox{\halign{\hfill#\hfill\cr Nice INLN 2006\#06 \cr}}
\hfill\vbox{\halign{\hfill#\hfill\cr\hglue-6cm December 2006\cr}}} \vglue2cm
\magnification=1150

\centerline{\bf Twin Paradox and Causality}
\bigskip
\bigskip
\bigskip
\bigskip 
\centerline{\rm T.\ Grandou $^{*,\diamond,\dagger}$\  and\  J.L. Rubin $^{*,\diamond}$}
\bigskip
\medskip 
\centerline{\it $^*$ Universit{\'e} de Nice-Sophia-Antipolis (UNSA), UFR Sciences,}
\centerline{\it $^\diamond$ Centre National de la Recherche Scientifique (CNRS), UMR6618,}
\centerline{\it $^\dagger$ Institut du Non Lin\'eaire de Nice UMR6618, 1361 route des Lucioles, 06560 Valbonne, France}
\centerline{e-mail: thierry.grandou@inln.cnrs.fr, jacques.rubin@inln.cnrs.fr}
\medskip\noindent
\bigskip\bigskip\medskip\bigskip\bigskip \centerline{\bf ABSTRACT}
\bigskip\smallskip\medskip After pointing out the historical avatar at the origin of a would be twin or clock paradox, we argue that, at least on a local scale, the (re-qualified) paradox is but a necessary consequence of the sole principle of causality.

\vfill  \eject
{\bf{I.  INTRODUCTION}}
\bigskip
As W.G. Unruh wrote more than 20 years ago, ``the
twin paradox has generated one of the longest standing controversies in
twentieth century physics"${}^{(1)}$. Of course, there is a big deal of literature on the subject, and a special merit, we think, is due to the contribution of Ref.2, which offers simple and synthetic enough a classification of the existing analyses.\medskip
 Indeed, if the mathematical concept of a (pseudo-) Riemannian manifold is 
adequate to the description of {\it{History}}, envisaged as
the set of all spacetime events, then, it must definitely be
stated that any sort of Langevin's twin paradox gets substantially re-interpreted. This is so because of a one, long known fact. With Sommerfeld's own words, ``As Minkowski once remarked to me, the element of 
{\it{proper-time}} is not an exact differential'' ${}^{(3)}$. Proper-time
lapses are therefore worldlines functionals. This statement ruins
the possible bases of any twin paradox, {\it{stricto sensu}}, and is a
general, geometric property of Riemannian and, to some extend, of
pseudo-Riemannian manifolds, not even restricted to Special or
General Relativity considerations.
\par For {compact} Riemannian manifolds, in effect, the
Hopf-Rinow theorem ensures that ${\cal{M}}$ is also
{\it{geodesically complete}}, and that any pair
of points can be joined by a geodesic of minimal length. If
two points are separated enough, they may be joined by another
geodesic whose length will therefore differ
from the minimal one. Differential aging
(of course, aging should properly be restricted to the pseudo-Riemannian case)
along intersecting worldlines will therefore arise as an intrinsic
geometrical property of the Riemannian manifold itself,
whatever the worldlines. In the physically interesting case of 
compact or so-called {\it{``t-complete"}} Lorentzian
manifolds, geodesic completness has been shown to extend to the  timelike and lightlike cases,
that are relevant to the twin paradox ${}^{(4)}$.
\medskip
This could be the end of the story. However, a geometrical
description is a final, {\it{effective}} description which,
when achieved, has erased the physical mechanisms at its
origin, providing us with the result, geometrical, of all
the {\it{forces}} having shaped it. A striking illustration of this claim can be learned out of the approach outlined in Ref.5 . That is, however elegant and satisfying in some
respects, such a description inevitably hides the fundamental,
irreducible physical principles at the origin of a given
phenomenon.

\medskip
In this letter, our intention is to somewhat {\it{unfold}} the kind of geometrical description alluded to above, so as to pin up the irreducible mechanism(s) and/or principle(s) responsible for the non-trivial differential aging phenomena. One may think, in effect, that present days technologies have shed new and decisive lights on questions that pure speculations revealed unable to fully elucidate. And it is conceivable that here is also one of the most compelling reason at the origin of a so long and vivid controversy. After all, "Science, has the same age as its technology", Wisdom says.\smallskip
Since a more detailed analysis will be proposed elsewhere, we will, here, alleviate the presentation and stay at the level of facts, principles and of their proposed articulations ${}^{(6)}$. To this end, a rapid review of the twin paradox is given in Section 2. Section 3 consists in a list of three concise points, two of them devoted to the historical and experimental developments of the matter, the third one, to the definition of proper-time lapses. In particular, a confusion at the origin of the famous paradox is put forth, and in order to preserve a long used terminology, the paradox appellation is accordingly re-defined. Within elementary Differential Geometry settings, the matter of Section 4 is the relation of the re-qualified twin paradox to the principle of causality. Then, Section 5 concludes the article.

\bigskip\bigskip\bigskip
{\bf{II. PARADOX SETTINGS}}\bigskip
As is well known, the famous paradox lies in its {\it{reciprocity}}. For either twin, in effect, non-trivial differential aging should manifest itself in exactly the same way, an obvious contradiction if the phenomenon is to be real. That is, the matter of {\it{reciprocity}} is to be examined with the one of {\it{reality}}. \smallskip Is this a sound consideration? A glance at History, even recent enough, should let no doubt about a positive answer ${}^{(7)}$. Major physicists, as well as philosophers, have long thought relativistic effects to be endowed with the status of {\it{appearances}} only, to fade away as soon as {\it{real}} comparisons of twins or clocks are duly achieved ${}^{(8)}$. And indeed, nothing in the special relativity formalism could prevent them to think so, quite on the contrary. From the onset, in effect, clocks and rods are assumed to be identical, in all of the inertial frames of reference~${}^{(9)}$.\smallskip In this respect, it is instructive to notice that even recent enough terminologies, like the one of {\it{parallax effects}} for {\it{time dilatation factors}}, for example, entail that connotation of {\it{appearances}}, though in a context where the relation {\it{real}}/{\it{apparent}} is clearly exposed ${}^{(10)}$.   \medskip
It would seem that the issue of {\it{reality}} should be disposed of easily. In effect, if special relativity parallax effects are nothing but pure appearances, then, the latter could simply fade away at the worldlines intersecting points : H. Bergson and H. Dingle are right (..to make a long story short!), there is neither any real differential aging, nor any paradox whatsoever. If, on the contrary, the so-called parallax effects are real, in the sense of persistent and actually measurable, then differential aging is real in the same acceptation, and an established {\it{reciprocity}} (symmetry) or {\it{non-reciprocity}} (asymmetry), would definitely have to be explained. \smallskip
 This will be the matter of the next section where it will be argued that this controversy should be looked upon as an {\it{avatar}} of the historical development of the special relativity formalisms. In the end, experiment had to decide. It did really, and it is the whole relation of the {\it{spacetime}} entity, to its many space {\it{and}} time coordinate realizations which had to be newly apprehended.

\bigskip\bigskip\bigskip
{\bf{III. PARALLAX AND PARADOX}}\bigskip

{\bf{A. A historical avatar}}\smallskip It is worth recalling how the idea of a paradox emerged and developed. As recently put forth in much details, standard special relativity formalisms are mixtures of Einstein's kinematics and Poincar\'e's group theory, that is, the group of {\it{scalar}} boosts in a given direction, which is a commutative one dimensional group.
\smallskip
This remark elucidates a large amount of the twin controversy, because it has long reduced the paradox discussions to an {\it{effective}} $1$ time + $1$ space dimensional case. There, as compared to the $1$ time + $2\  {\rm{or}}\ 3$ space dimension case to be discussed, the {\it{inertial frames reciprocity}} is so naturally preserved,  that it is not easy to figure out how an asymmetrical twin situation could possibly show up.\smallskip By {\it{inertial frames reciprocity}}, the following is meant. For any two given inertial frames of reference, $K(v)$ and $K'(v')$ (a third inertial laboratory frame, $K_0$, being understood with respect to which $v$ and $v'$ make sense), $K'(v')$ is seen from $K(v)$ the same, still opposite way, as $K(v)$ is seen from $K'(v')$, in agreement with the postulated equivalence of inertial frames~${}^{(11)}$. \smallskip 
Literature keeps track of that difficulty, which H. Bergson and H. Dingle, for example, could not think of another way than by appealing to {\it{appearances}}.  A part of truth was indeed contained in their point of view, the clue being that instead of a genuine paradoxical situation, to be re-defined shortly, the situation here is rather that of a parallax effect. Parallax effects are not Lorentz invariants, and may be thought of as {\it{appearances}}. But then, an important {\it{proviso}} should be made : these {{appearances}} give rise to persistent, really measurable effects, just like their $3$-euclidean space homologous do. \smallskip
Facts observed long ago in elementary particles accelerators are relevant to this situation. The example of $\pi$-meson beams can help fixing the idea and the terminology of the point being made. At a speed close to $c$, the $\pi$-mesons travel over distances corresponding to {\it{life-durations}} which can be a hundred times longer than their lifetimes of $10^{-8}$ sec. This is a physically measurable effect, and a persistent one since it is even used to keep the $\pi$-meson factory away from the experimental zone.\smallskip However, there is nothing more here than a pure parallax effect, due to the fact that the 1 time + 1 space laboratory axes are {\it{hyperbolically}} rotated with respect to the inertial beam spatio-temporal axes : the laboratory {\it{life-duration}} measurement is not performed in the $\pi$-meson beam {\it{proper-frame}}, and provides the real measurement of a real parallax effect ${}^{(3,12)}$. Under the exchange of inertial reference frames, the perfect symmetry of this parallax effect has led to the idea of a paradox when clocks (twins) had both to run fast with respect to the other!  
\smallskip 
Now, concerning Lorentz invariant quantities, there is no paradox nor any asymmetry. Proper-time lapses are the same in either cases, and a beam-embarked clock would deliver the same indication as a laboratory one, an averaged lifetime of $10^{-8}$ sec.
\smallskip 
Though sometimes misleading, Science does not often revoke its original terminologies, and we will keep using the word of "paradox". In the present case, however, it seems appropriate to reserve the paradoxical epithet for those situations involving invariant quantities only. In this way, any risk of confusion with parallax effects is avoided, right from the onset. In the more elaborate case of general relativity, a twin paradoxical situation will be translated into the terms of the present article's Introduction : two point-events of a given spacetime manifold can be joined by arcs corresponding to different proper-time intervals .. a still rather counter-intuitive fact indeed, and a one already present at the more elementary scale of Minkowski space geometry, as we will see. 

\bigskip
{\bf{B. A closely related and remarkable mechanism}}\smallskip
Other specific relativistic effects are worth of attention in themselves, as well as in their close relation to the non trivial differential aging phenomenon. Since only certain quadratic combinations of them form invariant quantities, relativistic theories do not really discriminate spatial from temporal coordinates. Their symmetrical role is clearly read off the most classical transformation {\it{formulae}} 
$$x'=\gamma(x-vt)\ ,\ \ \ y'=y\ ,\ \ \  z'=z\ ,\ \ \ ct'=\gamma(ct-{vx\over c})\eqno(1)$$At the level of spatial variables only, this symmetry might have helped anticipating the existence of some non trivial behaviours, and these are indeed the {\it{Thomas-Wigner rotations}}. 
\smallskip
At one more spatial dimensions in effect (2 instead of 1), the situation begins to change in many respects. Considered from the laboratory frame $K_0$, two successive boosts, $K_0\rightarrow K({\vec {v}})\rightarrow K'({\vec {v'}})$, may have non-collinear velocity vectors, and then, the $1$-space unescapable {\it{inertial frames reciprocity}} looks jeopardized because from $K_0$, $K'({\vec {v'}})$ is no longer seen the same, still opposite way, as $K_0$ is seen from $K'({\vec {v'}})$ ! \smallskip With the axes of $K({\vec {v}})$ taken parallel to those of $K_0$, and the axes of $K'({\vec {v'}})$, parallel to those of $K({\vec {v}})$, it is assumed that ${\vec {v}}$ is along the $Ox$ direction of $K_0$, and ${\vec {v'}}=d{\vec {v}}$, along the $Oy$ direction of $K({\vec {v}})$. Then, the $K_0$ versus $K'$ relative velocity, expressed in $K'$ and in $K_0$, points to different directions. An expression like
$${\rm{d}}\theta= {| {\rm{d}}{\vec {v}}_{Oy}|\over |{\vec {v}}_{Ox}|}\left(1-{\sqrt{1-{{\vec{v}}_{Ox}^2\over c^2}}}\right)\eqno(2)$$
is a typical textbook equation, which to first order, accounts for such an angular difference~${}^{(11)}$. This shows that, contrarily to the non-relativistic case, inertial frames parallelism is not a transitive relation in special relativity theory.
\medskip
In view of this, the simplest way to recover an essential {{inertial frames reciprocity}}, consists in stating that seen from $K_0$, the spatial axes of $K'({\vec {v'}})$ have {\it{rotated}} some angle ${\rm{d}}\theta$~ ${}^{(13)}$. 
\medskip
Now, an important point is that the Thomas-Wigner rotation also defines a fundamental connection between the invariant element of inertial frame {\it{proper-orientation}}, ${\rm{d}}\theta$, and the invariant element of {{proper-time}}, ${\rm{d}}\tau$. In the most general circumstances, corresponding to compositions of boosts along different directions, both {\it{non-exact}} differential forms are simply proportional
$$ {\rm{d}}\theta=\omega_{th}\  {\rm{d}}\tau\eqno(3)$$where ${\rm{d}}\tau$, is the element of proper-time of $K({\vec {v}})$ (or of $K'({\vec {v'}})$, at this order), and where, along a given worldline, $\omega_{th}$ is the instantaneous Thomas-Wigner rotation velocity itself.
\bigskip
From this ``spatial side" of the same relativistic properties (i.e., the non vanishing of differential forms ${\rm{d}}\theta$ and ${\rm{d}}\tau$), the character of reality is manifest, and bears on Lorentz invariant quantities. The following is meant. Right after Thomas subtle discovery, it has become possible to find out a missing factor of ${1/ 2}$ (and not simply corrections of order ${v^2/ c^2}$) reconciling some fine structure of alkalis doublets calculations, with the corresponding experimental results.\smallskip Since then, the {\it{Global Positioning System}} accurate technology has provided Thomas-Wigner rotations, with all the guarantees of soundness, persistence and reality, at any scale~${}^{(7)}$. Likewise, in Ref.14, new quantum mechanical phenomena are also presented in their relation to the Thomas-Wigner rotation. Indeed, the reason for such a universality is that Thomas-Wigner rotations are pure kinematical effects, and do not depend on the scale or dynamics of the situation considered.
\bigskip
{\bf{C. The proper-time line functional}}
\smallskip Now, it matters to specify how are defined the proper-time lapses. They will be defined by means of a {\it{stratagem}} proposed by A. Einstein. The stratagem consists in : "{\it{..imagining an infinity of inertial frames moving uniformly, relative to the laboratory frame, one of which instantaneously matching the velocity of the considered system, a twin, a clock, a particle..}}"~${}^{(3)}$. Considering thus ${\cal{C}}$, a worldline of the spacetime manifold ${\cal{M}}$, the proper-time lapse is the line functional
$$\Delta({\cal{C}};{\cal{R}})=\int{\rm{d}}\tau_{{}_{{\cal{C}}/{\cal{R}}}}(p)\ ,\ \ \ \ \forall p\in {\cal{C}}\eqno(4)$$
where ${\cal{R}}$ stands for some inertial "laboratory" frame of reference, and its associated {\it{time-orientation}}. ${\cal{R}}$ is of course arbitrary, but conveniently choosen in practice ${}^{(7)}$. Thanks to the above procedure, it is worth realizing that the proper-time line functional is a mathematically well defined quantity for any worldline ${\cal{C}}$. In particular, it is consistent, irrespective of the global spacetime manifold geometry. Gravitation/curvature, if any, must show up as an emergent or reconstructed effect, once admitted the equivalence principle (if gravitation is to be accounted for by general relativity, of course) ${}^{(15)}$. \smallskip  Considering ${\cal{C'}}$, another worldline with two points in common with ${\cal{C}}$, say $0$ and $\iota$, the functional of Eq.(4) will produce a differential aging result of
$$\Delta({\cal{C}},{\cal{C'}};{\cal{R}})=\int_0^\iota {\rm{d}}\tau_{{}_{{\cal{C}}/{\cal{R}}}} (p)-\int_0^\iota {\rm{d}}\tau_{{}_{{\cal{C'}}/{\cal{R}}}} (p')\equiv \delta T\eqno(5)$$
In the general case, this special relativity result is non vanishing, in view of the path and path $4$-velocity distribution dependences which differentiate ${\cal{C}}$ from ${\cal{C'}}$ (see section IV). It is proposed to account for any acceptable twin paradox and associated asymmetry, in the following sense. Exchanging ${\cal{C'}}$ and ${\cal{C}}$ just amounts, as it should, to change $\delta T$ into $-\delta T$. This is in contradistinction to the pure parallax effect of subsection A, where such an exchange, leaving the (non-Lorentz invariant) measure of {\it{life-duration}} unaffected, had motivated the original idea of a paradox.

\bigskip\bigskip\bigskip
{\bf{IV. TWIN PARADOX AND CAUSALITY}}\bigskip
The tight connection which, on a local scale at least, relates the non trivial differential aging phenomenon to the principle of causality is worth exploring. To do so, a first important result must be recalled. Then, a few and elementary geometrical tools will be introduced to help realizing that the whole (re-qualified) twin paradox is nothing but a selected aspect of a basic issue of differential geometry. That is, how are connected to each others, the different tangent spaces to a given manifold ?
\bigskip
{\bf{A. An important theorem ..}}\smallskip
 At any point $p$ of the spacetime manifold ${{M}}$, the tangent space, ${\cal{T}}_p{{M}}$, where the element ${\rm{d}}\tau(p)$ is evaluated, is assumed to be the {\it{vectorial}} Minkowski spacetime $I\!\!M$. This hypothesis is nothing but the {\it{principle of local Lorentz character}}, which benefits from an unquestionable support in experimental physics ${}^{(15)}$. This principle is a corner-stone of any further geometrical construction of the overall spacetime manifold ${\cal{M}}$, the general relativity theory being one such example. 
\smallskip
As well known, the squared element of proper time, ${\rm{d}}\tau^2$, is preserved, in particular, by the transformations of the inhomogeneous orthochronous Lorentz group, which includes possible space inversions, but excludes time reversal.\smallskip Now, a partial ordering can be defined on $M$, the {\it{affine}} Minkowski spacetime over the {\it{vectorial}} one, $I\!\!M$, which is to be understood the special relativity spacetime manifold. This partial ordering expresses the principle of {\it{causality}}, attached to both (i) the existence of a finite speed limit concerning information transfers, and to (ii) the existence of a global time-orientation of $M$. Writing $x<y$, if an {\it{event}} at $x$ can influence another {\it{event}} at $y$, we will write, following Ref.16,
$$x<y \Longleftrightarrow Q_0(y-x)\equiv (y_0-x_0)^2-({\vec{y}}-{\vec{x}})^2>0\ ,\ \ \ \   \& \ \ \ y_0-x_0>0\eqno(6)$$
Let $f$ be a function (not even assumed to be linear or continuous!), defining a one-to-one mapping of $M$ into itself. If $f$ and $f^{-1}$ preserve the partial ordering (6), in the very sense that
$$x<y \Longrightarrow f(x)<f(y)\ \ \&\ \  f^{-1}(x)<f^{-1}(y)\ ,\ \ \ \ \forall x, y \in M\eqno(7)$$then, $f$ is said to be a {\it{causal automorphism}} of $M$. \smallskip
Causal automorphisms of $M$ form a group, ${\cal{G}}$, which may be dubbed the {\it{causality group}} of $M$. Then, an important theorem states that, at $3$ spatial dimensions, the Minkowski space causality group ${\cal{G}}$, co\"{\i}ncides with the inhomogeneous orthochronous Lorentz group, augmented with dilatations of $M$ (multiplication by a scalar) ${}^{(16)}$.
\bigskip
{\bf{B.  .. and a few geometrical tools}}\smallskip
In general relativity, the {\it{lapse of time}} line functional reads
$$\Delta({\cal{C}})=\int_{{\cal{C}}} {\rm{d}}^4x {\sqrt{g_{\mu\nu}(x){\rm{d}}x^\mu{\rm{d}}x^\nu}}\eqno(8)$$
In this expression, one may notice the absence of reference to any arbitrary frame, inertial or not, like the one, ${\cal{R}}$, appearing in Equations (4) and (5). As compared to special relativity, this is the sign of a better achievement; this goal, in effect, was pursued at by Einstein. The {\it{Credo}} being that physics should not depend on {\it{coordinatizations}}, and should be encoded in intrinsic geometrical properties of the spacetime manifold itself.\smallskip
Now, special relativity also, can be formulated a geometrical way, free of reference frames, and interesting aspects can be learned out of this approach :
\smallskip
- Special relativity is established a particular case of general relativity theory, in continuity with it, and in agreement with an early statement of H. Weyl ${}^{(17)}$.
\smallskip
- Though the inertial observers spacetimes are Minkowskian and isomorphic,  they are different {\it{physical spaces}} attached each to a given 4-velocity vector $u$, at a given point $p$ ${}^{(1,17)}$. 
\smallskip
It matters to know, thus, how are these different spacetimes connected to each others, that is, in which relations stand their different space and time coordinate maps.
Let $I\!\!M_u$ and $I\!\!M_{u'}$ be two vectorial Minkowski spacetimes, time-oriented along $u, u'$, two 4-velocity vectors tangent to a given twin worldline ${\cal{C}}$, at points $p$ and $p'$ respectively.  As will be discussed further on [after Eq.(19)], a {\it{natural correspondence}} between the spaces $I\!\!M_u$ and $I\!\!M_{u'}$ is provided by the pure (without rotation) Lorentz boost from $u$ to $u'$, and may be given the coordinate-independent (i.e., geometrical) expression of ${}^{(17)}$
$$ I\!\!B(u',u)=1-{(u'+u)\otimes (u'+u)\over 1+u'\cdot u}+2u'\otimes u\eqno(9)$$ The Lorentz scalar product is given by the metric, $g_0=diag\ (+1,-1,-1,-1)$, and the symbol $u'\otimes u$ stands for the linear mapping of $I\!\!M$ into itself,
$$\forall x, u, u'\in I\!\!M\ ,\ \ \ u'\otimes u \ : x\longmapsto u'(u\cdot x)\eqno(10)$$One has    $I\!\!B(u',u)u=u'$ and $I\!\!B(u',u)\circ I\!\!B(u,u')=I\!\!I$. The correspondence so established allows us to compare vectors belonging to different inertial spaces, and to define a notion of {\it{physical equality}} (and {\it{parallelism}}) of vectors, which is reflexive, but is not transitive. Referred to the coordinate axes of a given inertial space, those of the "stayed home twin" for instance, the Thomas-Wigner rotation is an expression of this non-transitivity.
\smallskip
In this respect, a peculiar feature is noteworthy. For any 3 non-coplanar 4-velocity vectors, $u, u'$ and $u''$ (equivalent to 2 non-collinear relative 3-velocities), the composition of 3 successive boosts, without rotation, is not a boost without rotation .. but a rotation without boost ${}^{(17)}$ ! 
$$I\!\!B(u,u'')\circ I\!\!B(u'',u')\circ I\!\!B(u',u)=I\!\!R_u(u',u'')\eqno(11)$$
Passing from the three vectors, $u, u'$ and $u''$,  to the two relative velocities, ${\vec {v}}_{Ox}$ and ${\rm{d}}{\vec {v}}_{Oy}$, the above expression just reproduces the Thomas-Wigner Rotation of Eq.(2).  
\bigskip
Now, causality "lives upstairs" of those {\it{vectorial}}-referred objects, at the level of the {\it{affine}}, base-pointed Minkowski spaces, $M_o$, $M_p$, .., $M_{\iota}$. Their elements are not vectors, but point-events, and their labels refer, here, to the points of a given timelike worldline, ${\cal{C}}$.
\bigskip
{\bf{C. Application to the twins}}\smallskip 
 An {\it{absolute}} Minkowski spacetime of reference, $M$, must be choosen. For the sake of twin paradox, $M$ can be taken as being the inertial stayed home twin space with, by definition, the corresponding 4-velocity $u_0$. Within some geometrical terminology in use, $M$-time is $u_0$-time, and $M$-space, the $u_0$-space ${}^{(17)}$. The spacetime $M$ is thus the point $O$-referred affine Minkowski space over $I\!\!M_{u_0}$. It is endowed with a distance $d_0$, defined through the Lorentzian non degenerate quadratic form $Q_0$ of Eq.(6),
  $$\forall p, p'\in M\ ,\ \ \ \ d_0(p,p')=Q_0(p-p')\eqno(12)$$ 
\smallskip
The travelling twin history is accounted for by a twice-differentiable mapping $r(s)$ of an interval $[s_i,s_f]$ into $M$, whose range is the twin's worldline ${\cal{C}}=\{r(s)\ |\ s\in [s_i,s_f]\subset I\!\!R\}$. With $r(s_i)=O$, the mapping $r(s)$ satisfies the relations
$$\forall s\in [s_i,s_f]\ , \ \  {{\rm {d}}r(s) \over {\rm {d}}s}\equiv \dot r(s)\ ,\ \ \  {{\dot r}^2(s)}\equiv{\dot r(s)}\!\cdot\! {\dot r(s)} \equiv g_0(\dot r(s),\dot r(s))=Q_0(\dot r(s))=1\eqno(13)$$  as well as the usual frame-independent relation of $M$-time $t$ to the worldline ${\cal{C}}$ proper-time, $s$
$$   u_0\cdot {\dot {r}}(s)\  {\rm {d}}t={{\rm {d}}s }\eqno(14)$$
Besides $r(s_i)=O$, there is no loss of generality in completing the ``initial data'' with the condition ${\dot r}(s_i)=u_0$. This condition just corresponds to a given twin experimental protocol while giving rise to more beautiful equations (in particular, {\it {fonctorial}} relations are made more transparent in this way ${}^{(6)}$).  
\medskip
At any point $p\in {\cal{C}}\subset M$, let ${\cal{T}}_pM$ be the space  tangent to $M$ at $p$. From $(M, d_0)$, the space ${\cal{T}}_pM$ inherits a vectorial Minkowski structure 
$$\forall u, u' \in {\cal{T}}_pM\ ,\ \ \ \ g_p(u,u')=g_0(u,u')={1\over 2}\left(Q_0(u+u')-Q_0(u)-Q_0(u')\right)\eqno(15)$$
One can denote $I\!\!M_p$ the set $({\cal{T}}_pM, g_p)$ 
, and to $I\!\!M_p$ is trivially associated the affine Minkowski spacetime $M_p$, with the same causality group, ${\cal{G}}$, as $M$.  This applies, of course, to any point $r(s)$ of the travelling twin worldline, ${\cal{C}}$. 
 
\smallskip
But in this latter situation, because $\dot r(s)$ is, according to (13), a $4$-velocity vector for all $s$, it can be shown that $M_{r(s_i)}$- and $M_{r(s)}$- causalities are the same. That is, between the affine Minkowski spacetimes $M_{r(s_i)}$ and $M_{r(s)}$, for all $s\in [s_i,s_f]$, there exists a causal isomorphism $\varphi_{s,s_i}:M_{r(s_i)}\rightarrow M_{r(s)}$ sending the $M_{r(s_i)}$ base-point, ${r(s_i)}=O$, onto the base-point of $M_{r(s)}$, and mapping the ($O$- and ${\dot r}(s_i)$ -referred) partial ordering (6) of $M_{r(s_i)}$, onto the ($r(s)$- and $\dot r(s)$ -referred) partial ordering (6) of $M_{r(s)}$. One has simply,
$$\forall q\in M_{r(s_i)}\ ,\ \ \ \ 
 \varphi_{s,s_i}(q)=r(s)+I\!\!L_{s,s_i}\left(q-{r(s_i)}\right)\ \in M_{r(s)}\eqno(16)$$
where, in virtue of the theorem of subsection A, $I\!\!L_{s,s_i}$ is an element of ${\cal{L}}_+^\uparrow$, the homogeneous orthochronous Lorentz group, or a dilatation, which, both, act on vectors of $I\!\!M_{\dot r(s_i)}(\equiv I\!\!M)$. Equation (16) makes it clear that the mapping $\varphi_{s,s_i}$ can be called ``the affine application {\it{over}} the linear application $I\!\!L_{s,s_i}$'' (by the way, since $I\!\!L_{s,s_i}$ defines a vectorial space isomorphism, $\varphi_{s,s_i}$ is also a causal diffeomorphism, in view of the {\it{Inverse mapping theorem}}). 
  \medskip
At this point, because relativistic theories do not discriminate between spatial and temporal coordinates, it matters to make contact with the closely related mechanism of subsection~{{III.B}}.  At base-point $r(s_i)=O$, one can attach a {\it{tetrad}} of orthonormal basis vectors spanning the vectorial Minkowski space $I\!\!M_{{\dot r}(s_i)}=I\!\!M$, the set $\{e_0(s_i)\equiv {\dot r}(s_i),\  e_j(s_i)\ ;\  j=1,2,3 \}$. Since the affine Minkowski spacetimes $M_{r(s_i)}$ and $M_{r(s)}$ are mapped into each other by the causal isomorphism $\varphi_{s,s_i}$, the tetrad $\{ {{\varphi_{s,s_i}}_\ast}\   e_0(s_i),\  {{\varphi_{s,s_i}}_\ast}\  e_j(s_i)\ ;\  j=1,2,3 \}$ will span the vectorial Minkowski space $I\!\!M_{\dot r(s)}$, where, in the standard definitions of differential geometry, ${{\varphi_{s,s_i}}_\ast}$ is the differential of the application $\varphi_{s,s_i}$, at point $r(s_i)$. Of course, then,
  $${{\varphi_{s,s_i}}_\ast}=I\!\!L_{s,s_i}\eqno(17)$$ 
 \smallskip 
 Here, a particular case is of interest in relation to the stratagem of subsection {{III.C}}. In effect, the stratagem requires that at any proper instant $s$ of the travelling twin worldline, the co-moving tetrad complies with the identity $e_0(s)\equiv \dot r(s)$. The other three spacelike vectors, $e_j(s), j=1,2,3$, can be physically realized as gyroscopes, and formally thought of as {\it{gyrovectors}} ${}^{(18)}$. In this case, the tetrad is said to be {\it{Fermi-Walker transported}} along ${\cal{C}}$, and the tangent mapping ${{\varphi_{s,s_i}}_\ast}$ results of a composition of an infinite series of infinitesimal boosts ranging from proper-instants $s_i$ to $s$, along ${\cal{C}}$
 $$\eqalignno{
I\!\!L_{s,s_i} &= I\!\!B\left({{\dot r}(s)},{{\dot r}(s-{\rm{d}}s)}\right)\circ\ ..\  \circ\ I\!\!B\left({{\dot r}(s_i+{\rm{d}}s)},{{\dot r}(s_i)}\right)\cr
&= \prod_{s'=s_i}^s\ \circ\ \left(\ I\!\!I + {\rm{d}}s'\ ({\dot r}(s')\wedge {\ddot r}(s'))\ \right)&(18)\cr}$$ 
where the symbol $a\wedge b$ is introduced as a shortand, still standard notation for the antisymmetric product
$$\forall z\in I\!\!M,\ \ \ ({\dot r}\wedge {\ddot r})z={\dot r}\left({\ddot r}\cdot z\right)-{\ddot r}\left({\dot r}\cdot z\right)\eqno(19)$$  
    \smallskip
  One may observe that the neighbouring tangent spacetimes, the $I\!\!M_{\dot r(s)}$, are ``connected" to each other by pure boosts, without rotation (9). For a number of authors though, this correspondence has long been recognized to entail an irreducible part of convention ${}^{(19)}$. This is because it relies on a peculiar convention of synchronization, namely, the Einstein's convention, when many others {\it{look}} possible ${}^{(2,17)}$. But on the other side, a sound argument has recently been proposed, emphasizing how  {\it{natural}} and inherent to the relativity theory the Einstein's convention is : to be consistent, a change of convention should only be thought of within a full modification or a {\it{deformation}} of the relativity theory itself ${}^{(20)}$.  
  \medskip
  Now, to be read in $I\!\!M_{{\dot r}(s_i)}$, a vector of $I\!\!M_{{\dot r}(s)}$ must undergo the pure boost transformation of $I\!\!B({{\dot r}(s_i)},{{\dot r}(s)})$ ${}^{(17)}$. This gives rise to $M_{r(s_i)}$ causal automorphisms, elements of ${\cal{G}}$, the $M$-causality group, such as
 $$f_{s,s_i} : M\rightarrow M ,\ \ \  \ \ \ f_{s,s_i}(q)=r(s)+I\!\!B\left({{\dot r}(s_i)},{{\dot r}(s)}\right)\circ I\!\!L_{s,s_i}\left(q-r(s_i)\right)\eqno(20)$$

  \smallskip
  Then, the automorphism causal series, $\{f_{s,s_i} \ ; s\in [s_i,s_f]\}$, entails the travelling twin worldline,
  $$\{f_{s,s_i}\left(r(s_i)\right)/\ s\in [s_i,s_f]\}=\{r(s)/\ s\in [s_i,s_f]\}={\cal{C}}\eqno(21)$$
 the instantaneous relation of $M$-time to $M_{ r(s)}$-time
   $${{\rm{d}}s }={\dot r}(s_i)\cdot {f_{s,s_i}}_\ast\left({\dot r}(s_i)\right)\ {\rm{d}}t\eqno(22)$$
 as well as the instantaneous Thomas-Wigner rotation of the spatial coordinate axes, the $e_j(s)$, with respect to the stayed home twin axes, provided that one has $e_0(s)={\dot r}(s_i)$ (this condition, in effect, is mandatory in order to have identical $3$-dimensional spaces, and define meaningful rotations ${}^{(17)}$),
 $$e_j(s)={{f_{s,s_i}}_\ast}\left(e_j(s_i)\right)\ ,\ \ \ \  s\in [s_i,s_f]\ ,\ \ \ \  j= 1,2,3.\eqno(23)$$
  
In the end, one and the same causal series, $\{f_{s,s_i}\ ;\  s\in [s_i,s_f]\}$ encodes not only the continuous timelike curve ${\cal{C}}$ itself (21), but also all of its local (instantaneous) and global (integrated) characteristics:

\smallskip - The fact that the twin's spaces may be found rotated with respect to each other, even though no torque has been met during the trip (23).

- The non-trivial differential aging phenomenon, by integration along ${\cal{C}}$ of the non-exact differential proper-time 1-form. Effectively, an equality of proper-time lapses, $s_f-s_i$ and $ t_f-t_i $ does not hold in the general case, where one has, if again ${\dot r}(s_f)={\dot r}(s_i)$, 
$$t_f-t_i={\int\!\!\!\!\!{\cal{C}}}_{s_i}^{s_f} {\rm{d}}s\ \left({\dot r}(s_i)\cdot {{f_{s,s_i}}_\ast}\left({\dot r}(s_i)\right)\right)^{-1}\ \leq\  s_f-s_i\eqno(24)$$
for those of the tangent automorphisms, ${{f_{s,s_i}}_\ast}$ that are in ${\cal{L}}_+^\uparrow$. If, instead, ${{f_{s,s_i}}_\ast}$ is a {\it{contraction}}, i.e., a global dilatation by a factor smaller than 1, then the opposite relation obviously results, of $t_f-t_i> s_f-s_i$, indicating that causality alone does not tell whom, of either twin, is aging faster. \smallskip Note that equation (24) is made more familiar if we keep in mind that the term of $\left({\dot r}(s_i)\cdot {{f_{s,s_i}}_\ast}\left({\dot r}(s_i)\right)\right)^{-1}$ is ordinarily thought of as the usual $\gamma^{-1}$ factor of ${\sqrt{1-{\vec{v}}^2(s)/c^2}}\leq 1$. This sends us back to the controversial {\it{paradigm}} of inertial frames of reference and their relative uniform velocities. However, in the present geometrical context, it matters to realize that (24) is not bound to that interpretation and/or derivation. Rather, it is worth emphasizing that it is again a pure consequence of causality, without it being necessary to call for anything else. This can be phrased as follows. Causality is implemented on $M$ through the partial ordering relation (6), that is, through the Lorentzian non-degenerate quadratic form $Q_0$. Now, because ${\dot r}(s_i)$ is timelike, the inverted Cauchy-Schwarz inequality holds ${}^{(21)}$  
$$\forall v\in I\!\! M\ ,\ \ \ \ \ {\dot r}^2(s_i)\ v^2\ \leq\  ({\dot r}(s_i)\cdot v)^2\eqno(25)$$where there is equality whenever $v$ and ${\dot r}(s_i)$ are linearly dependent. Furthermore, since ${\dot r}(s_i)$ is of unit length (${\dot r}^2(s_i)=1$), for those ${{f_{s,s_i}}_\ast}$ that are in ${\cal{L}}_+^\uparrow$, one has $\ {\dot r}(s_i)\cdot {{f_{s,s_i}}_\ast}\left({\dot r}(s_i)\right)\geq 1$, and thus the inequality (24).

\bigskip\bigskip\bigskip
{\bf{V. CONCLUSION}}\bigskip
Since it was launched by P. Langevin in 1911 (and was indeed explicit in the Einstein's 1905 famous article), the twin paradox proper framework was soon identified with general relativity theory, because of accelerations to be considered along a twin worldline, at least. This point of view was adopted by Einstein and supported by M. Planck. However, it has long been recognized to be at fault, for both theoretical and experimental reasons ${}^{(22)}$. In particular, accelerations can be consistently dealt with in flat spacetime manifolds, and should no way be mistaken for gravitation ${}^{(15)}$. Moreover, the argument based on accelerations can also be circumvented, so as to bring the paradox back to its original special relativity birthplace ${}^{(23)}$.\smallskip 
 In this article, our intention has been to look for the principle at the origin of a so counter-intuitive, but established fact as ``the non-trivial differential aging phenomenon''. And to this end, it was certainly appropriate to look for such a principle in the simpler structure where the phenomenon is manifest, that is, over the local scale of a Minkowski spacetime manifold.
 \smallskip 
 If gravity, as described in general relativity, is responsible for another source of non-trivial differential aging contributing on the same footing as special relativity effects in some situations, a would be twin paradoxical case, in its conventional acceptation at least, is not a natural issue of the general relativity framework ${}^{(24)}$. Two essential reasons may be proposed.
\smallskip - First, as we have seen, the conventional twin paradox should be considered an {\it{avatar}} of the {{hybrid}} nature of special relativity standard formalisms, where the paradox was first conceived, discussed and confused with the (full reciprocity of) special relativity (non-invariant) perspective effects. As such, it has no natural expression in the general relativity formalism. A flavour of this can be grasped out of the twin paradox resolution proposed, in this context, by H. Reichenbach ${}^{(25)}$. 
 \smallskip - Then, when such an elaborated state of affairs as general relativity theory is reached, any possible twin paradox content gets translated into the pseudo-Riemannian theorems mentioned in the introduction; that is, in terms of the geometrical properties of the spacetime manifold~${\cal{M}}$. \medskip
  
And indeed, the same explanation is proposed here for the twin paradox, at the more local scale of a Minkowski spacetime manifold, $M$. \smallskip
The twin paradox, in effect, has first been re-qualified into the path-functional dependence of proper-time lapses, while preserving the name, so as to keep in touch with the historical terminology. The paths are continuous timelike curves which, in virtue of a famous theorem, completely ``encodes'' the geometrical, differential and topological structures of $M$ ${}^{(26)}$. This is why the ``the non-trivial differential aging phenomenon'' can, likewise, be thought of as a property of the Minkowski spacetime geometry.

 \medskip
 
 In the end, causality revealed to be the principle, and the only one, from which the (re-qualified) twin paradox and the somewhat correlated Thomas-Wigner rotations come from. \smallskip Ten years ago, causality was advocated to provide a global constraint on the possible twins histories, labelled, each, by some experimental/theoretical synchronization device ${}^{(2)}$.\smallskip Now, more than an overall constraint on the twin's histories, one can see here, how the sole principle of causality stands at the very source of the twin paradox. That is, how preservation of causality along continuous timelike worldlines necessarily involves a functional dependence of proper-time lapses on the paths themselves. \medskip
 
 That space and time should be considered as melt into a one and single spacetime entity is, definitely, a most salient feature of relativity theories. That this necessity comes from the need of providing causality with a sound enough support is, we think, a remarkable fact. It would seem to point to the requirement for {\it{History}} to be meaningful, in a physical and thus restricted, still crucial sense.\smallskip
 Beyond the twin paradox itself, one may remark that it is possible to derive the whole special relativity theory out of a single and intuitively clear principle of causality. In this respect, the famous paradox may be looked upon in analogy with those situations encountered in Mathematics, where unquestionable axioms are able to generate counter-intuitive .. if not ``paradoxical" consequences ${}^{(27)}$. 

 \bigskip \bigskip \bigskip\vfill\eject
 {\bf{Acknowledgement}}
  \bigskip
  One of us (T.G.), thanks J.M. Levy-Leblond for his comments and for having suggested this analysis, and M. Le Bellac for several critical readings. We also thank G. Rousseaux who provided us with many interesting references on the subject.

 \bigskip\bigskip\bigskip
{\bf{References}}\bigskip
 1.  W.G. Unruh,  {\it Am. J. Phys.} {\bf{49}}(6), 589 (1981).
 \smallskip
 2.  T. A. Debs,  and M. L. G. Redhead, {\it Am. J. Phys.} {\bf{64}}(4) 384 (1996).\vskip0.05truecm $\ \ \  $ D. Malament, {\it No{\^u}s} {\bf{7}}, 293 (1977). 
 \smallskip
 3.  Y. Pierseaux, {\it{ Annales de la Fondation Louis de Broglie}} {\bf{29}},  57 (2004).
 \smallskip
 4.  A. Avez, {\it Annales de l'Institut Fourier}, {\bf{13}}(2), 105 (1963).
\smallskip
 5.  S. L. Adler, {\it Rev. Mod. Phys.} {\bf{54}}(3) 729 (1982).
 \smallskip
 6.  T. Grandou, and J. L. Rubin,   Work in completion.
 \smallskip
 7.  N. Ashby,  {\it Living Rev. Relativity} {\bf{6}}(1), 18 (2003)\vskip0.05truecm $\ \ \ \ $
[Online Article]: cited on 28 january 2003,\vskip0.05truecm $\ \ \ \ $
http://www.livingreviews.org/Articles/Volume6/2003-1ashby/
 \smallskip
 8.  H. Bergson, {\it Duration and Simultaneity} (Clinamen Press, Manchester, 1999)
 \vskip0.05truecm $\ \ \ \ $ H. Dingle, {\it Bulletin of The Institute of Physics}, {\bf{7}},  314 (1956). 
 \smallskip
 9.  See Ref.3, for example, pp. 72, 82 and 88, note 13.
 \smallskip
10. J-M. Levy-Leblond, {\it La Recherche}, {\bf{96}}(10), 30 (1979).
 \smallskip
11. R. M. Eisberg, {\it Fundamentals of Modern Physics} (J. Wiley, 1967). \vskip0.05truecm $\ \ \ \ $   J. Aharoni,  {\it The Special Theory of Relativity} (Dover Publications, 1965).
 \smallskip
12. J. P. P\'erez, {\it{Relativit\'e}} (Dunod, Paris, 1999).
  \smallskip
13. K. R. MacKenzie, {\it Am. J. Phys.} {\bf{40}}(11), 1661 (1972).
 \smallskip
14. P.  L\'evay,  {\it{J. Phys. A: Math. Gen.}}   {\bf{37}},  4593 (2004).
 \smallskip
15. C. W. Misner,  K. S. Thorne, and J. A. Wheeler,\vskip0.05truecm $\ \ \ \ $ {\it Gravitation} (W. H. Freeman, San Francisco, 1973), p.164.
 \smallskip
16. E. C. Zeeman, {\it{J. Math. Phys.}} {\bf{4}},  490 (1963).
  \smallskip
17. H. Bacry, {\it{Le\c cons sur la Th\'eorie des Groupes}} (Gordon \& Breach, 1967);
\vskip0.05truecm $\ \ \ \ $ T. Matolcsi and A. Goher, {\it{Stud. in Hist. and Phil. of Mod. Phys.}} {\bf{32}} (1) 83 (2001). \vskip0.05truecm $\ \ \ \ $ H. Weyl, {\it{Space, Time, Matter}} (Dover Publications, Inc., 1952). 
 \smallskip
 18. A. A. Ungar, {\it{Fundamental Theories of  Physics}} {\bf{117}}\vskip0.05truecm $\ \ \ \ $ (Kluwer Academic Publishers Group, 2001).
  \smallskip
19. See, for example, in Ref.17, H. Bacry  p.261.
  \smallskip
20. D. Malament,(2006). {\it{Classical Relativity Theory}} in {\it{Handbook of the Philosophy of\vskip0.05truecm $\ \ \ \ $ Science. Volume 2: Philosophy of Physics}} (D.M. Gabbay, P. Thagard and J.Woods,\vskip0.05truecm $\ \ \ \ $ Editors, Elsevier, 2006).
 \smallskip
21. D. Giulini, {\it{Algebraic and geometric structures of Special Relativity}},\vskip0.05truecm $\ \ \ \ $arXiv:math-ph/0602018v1, 2005.
 \smallskip
22. C. S. Unnikrishnan, {\it{Current Science}} {\bf{89}} (12), 2009 (2005).
  \smallskip  
23. H. Bondi, {\it{Discovery}}, {\bf{18}}, 505 (1957).
 \smallskip
24. J. C. Hafele, and R. E. Keating, {\it{SCIENCE}} {\bf{177}}, 166 (1972).
    \smallskip  
25. H. Reichenbach, {\it{Boston Studies in the Philosophy of Science}}, \vskip0.05truecm $\ \ \ \ $ {\bf{XXII}}   (D. Reidel Publishing Company, 1976), pp. 447-449.  
  \smallskip
26. D. Malament, {\it{J. Math. Phys.}} {\bf{18}}, 1399  (1977).
   \smallskip
27. P. de la Harpe, {\it{Panoramas \& Synth\`eses}} {\bf{18}},  39 (2004).

\end